\documentclass[10pt,letterpaper]{article}
\usepackage[top=0.85in,left=2.75in,footskip=0.75in]{geometry}

\usepackage{amsmath,amssymb}

\usepackage{changepage}

\usepackage[utf8x]{inputenc}

\usepackage{textcomp,marvosym}

\usepackage{cite}

\usepackage{nameref,hyperref}

\usepackage[right]{lineno}

\usepackage{microtype}
\DisableLigatures[f]{encoding = *, family = * }

\usepackage[table]{xcolor}

\usepackage{array}

\newcolumntype{+}{!{\vrule width 2pt}}

\newlength\savedwidth

\raggedright
\usepackage[aboveskip=1pt,labelfont=bf,labelsep=period,justification=raggedright,singlelinecheck=off]{caption}

\makeatletter
\renewcommand{\@biblabel}[1]{\quad#1.}
\makeatother

\date{}

\usepackage{lastpage,fancyhdr,graphicx}
\usepackage{epstopdf}
\fancyheadoffset[L]{2.25in}
\fancyfootoffset[L]{2.25in}

\begin{document}
\vspace*{0.2in}

\begin{flushleft}
{\Large
\textbf\newline{EduCTX: A blockchain-based higher education credit platform} 
}
\newline
\\
Muhamed Turkanovi\'{c}\textsuperscript{*},
Marko H{\"o}lbl, 
Kristjan Ko\v{s}i\v{c}, 
Marjan Heri\v{c}ko, 
Aida Kami\v{s}ali\'{c} 

\bigskip
 University of Maribor, Faculty of Electrical Engineering and Computer Science, \\ Koro\v{s}ka cesta 46, 2000 Maribor, Slovenia
\\
\bigskip

%
%





* muhamed.turkanovic@um.si

\end{flushleft}
\section*{Abstract}
Blockchain technology enables the creation of a decentralized environment where transactions and data are not under the control of any third party organization. Any transaction ever completed is recorded in a public ledger in a verifiable and permanent way. Based on blockchain technology, we propose a global higher education credit platform, named EduCTX. This platform is based on the concept of the European Credit Transfer and Accumulation System (ECTS). It constitutes a globally trusted, decentralized higher education credit and grading system that can offer a globally unified viewpoint for students and higher education institutions (HEIs), as well as for other potential stakeholders such as companies, institutions and organizations. 
As a proof of concept, we present a prototype implementation of the environment, based on the open-source Ark Blockchain Platform. Based on a globally distributed peer-to-peer network, EduCTX will process, manage and control ECTX tokens, which represent credits that students gain for completed courses such as ECTS. HEIs are the peers of the blockchain network. The platform is a first step towards a more transparent and technologically advanced form of higher education systems. 
The EduCTX platform represents the basis of the EduCTX initiative which anticipates that various HEIs would join forces in order to create a globally efficient, simplified and ubiquitous environment in order to avoid language and administrative barriers. Therefore we invite and encourage HEIs to join the EduCTX initiative and the EduCTX blockchain network.



\section*{Introduction}

Based on the concept of the European Credit Transfer and Accumulation System (ECTS) we propose a global blockchain-based higher education credit platform, named EduCTX. The proposed system will exploit the benefits of the blockchain, as a decentralized architecture, offering security, anonymity, longevity, integrity, transparency, immutability and global ecosystem simplification, in order to create a globally trusted higher education credit and grading system. As a proof of concept we will present a prototype implementation of the platform, which is based on the open-source Ark Blockchain Platform \cite{2016All-in-OneSolutions}. 

The scientific contribution is to provide a distributed and interoperable architecture model for the higher education credit system which addresses a globally unified viewpoint for students and institutions. Potential employers can benefit from the proposed system. 

Students can take advantage of having their completed course history in a single and transparent view, as well as universities which have this data accessible and up to date, regardless of a student's educational origins. On the other hand, different organizations (such as employers, universities, etc.) as potential users of the system, can validate the provided information after a student's permission is obtained. 

The structure of the paper is as follows. The current section presents in detail the motivation and the contributions of the paper. Related works present important projects and research in the field related to this study, while the Background section introduces the background and preliminaries of the work. The main contribution of the paper is presented in the Proposed EduCTX Platform and Prototype implementation sections. The section Proposed EduCTX Platform will present in detail the proposed concept of the platform, while Prototype implementation covers the technical parts, i.e. implementation, functioning, real-life example, etc. Some reflections and issues of the work are described in detail in the Discussion section. Finally, the Conclusion and Future work provides a summary of the proposed solution and some future plans.

\subsection*{Motivation}
\label{sec_motivation}

The majority of higher education institutions (HEIs) keep their students' completed course records in proprietary formats. These databases are structured to be exclusively accessed by an institution's staff and in dedicated online systems, hence with little or no interoperability. Furthermore, the majority of institutions have their own specialized system for keeping students' completed course records, which preserves the proprietary data structure of the database. In general, these databases are hosted in a data center inside the HEI, with restricted access to its IT professionals. Students can have external access to their data in a restricted, password protected manner, only to view or print their completed course records (some systems enable and log students' online check-in and check-out exams). There are several vital points in regards to such systems, including standardization of data, storage location, safety and how to filter, analyze and securely share such data. Connected with these issues, HEIs maintain the students' completed course records indefinitely. This is required for legal reasons, depending on a country's policy. Also in the majority of cases, institutions do not share their students' data, not even the completed course records. Hence, students can experience difficulties transferring to another HEI, while still preserving and proving their completeness of courses from the previous institution. This problem is even more vivid in cases when a student wants to transfer to another country, where a language, script and administrative barrier exist. Moreover, these records are usually stored in different standards, which make it difficult to exchange records between HEIs.

In cases when the student applies for a job position and has to prove his/her academic degree in a foreign country, problems arise from the centralized storage of students' complete course records due to their inaccessibility, lack of standardization, etc. The students have to translate and nostrificate their academic certificate which can be a complex and time-consuming process. The nostrification process includes the translation of all official documentation into the language of the host institution, which has to review and validate every aspect of the documentation in order to examine matching or diverging content. 

Furthermore, after completing their education, students sometimes have no access to the online academic grading system. In such a case, if a student loses his/her academic certificates, he/she needs to visit their home HEI and request a new copy, which can be a costly and time consuming process.       

Although there are some consolidated standards for the academic credit system like the ECTS, the adoption and implementation of a global decentralized, trusted, secure credit platform, is a challenge. Many of the obstacles come from the fact that students' academic records are sensitive and have complex management regulations in place.

\subsection*{Contribution}
\label{sec_contribution}

We propose a blockchain-based decentralized higher education credit platform, named EduCTX. It builds on the distributed peer-to-peer (P2P) network system. These systems are flexible, secure and resilient because of their storage capacity and resource sharing on a global scale \cite{Coulouris2011DistributedEdition}. The EduCTX platform transfers the higher education credit and grading system from the analog and physical world into a globally efficient, simplified, ubiquitous version, based on the blockchain technology. The EduCTX platform is the basis of our EduCTX initiative (more information available at: eductx.org), which envisions a unified, simplified and globally ubiquitous higher education credit and grading system. Through the initiative we plan to further on advance and develop the EduCTX concept.  

\section*{Related Works}
\label{sec_rw}

Blockchain technology aims at creating a decentralized environment where no third party is in control of the transactions and data\cite{Yli-Huumo2016WhereReview}. It is used in several domains due to its benefits in distributed data storage and the possibility of audit trails.

In healthcare, several approaches have been introduced in the field of electronic health records (EHR) \cite{Alex2017OmniPHR:Records, Kemkar2012FormulationConcept, He2013, Shae2017, Xia2017MeDShare:Blockchain, Mettler2016BlockchainHere, Azaria2016}. Clinical trials or data access and permission management are fields in which the technology can be applied \cite{Shae2017, Azaria2016}. Closely related to EHRs is interoperability, where blockchains have also been employed \cite{He2013, Xia2017MeDShare:Blockchain}. Some authors even claim that it could revolutionize healthcare \cite{Mettler2016BlockchainHere}. They give examples as smart and public healthcare management, which can benefit the patient, by using blockchain technology to fight counterfeit drugs in the pharmaceutical industry.

However, healthcare is only one of the possible blockchain application domains. Due to the transparency of the technology, the domains of government and business also try to apply the technology and harvest its benefits \cite{Hou2017, lnes2017BlockchainE-Government, lnes2016BeyondTechnology, Morabito2017BlockchainGovernance, Qi2017Blockchain-PoweredE-Democracy}. The blockchain is applied in e-government scenarios \cite{Hou2017}, smart government \cite{lnes2016BeyondTechnology}, etc. In the business domain new concepts and systems arise (e.g. electronic cash systems, business processes, etc.) \cite{Sidhu2017, 2016InternetApplications, Rimba2017ComparingExecution}.

Even in logistics and transportation blockchain technology can be applied \cite{Yuan2016, 8010820}. In this way, new intelligent transportation systems are developed. Additionally, energy production, management and trading can foster the benefits of the blockchain \cite{Imbault2017, ZhumabekulyAitzhan2016SecurityStreams, Mengelkamp2017AMarkets}. Smart grids and different smart technologies can employ the technology to optimize their operation \cite{Mengelkamp2017AMarkets} and new business opportunities can be developed \cite{ZhumabekulyAitzhan2016SecurityStreams}.

Even in the emerging field of the Internet of Things (IoT), blockchain technology can be used in different scenarios and forms \cite{Conoscenti2017, Qi2017Blockchain-PoweredE-Democracy, Zhang2017TheThings, Ouaddah2017TowardsIoT, 2016InternetApplications, Kshetri2017CanThings, Dorri2017BlockchainHome}. These include the management of privacy and security of IoT \cite{Dorri2017BlockchainHome, Ouaddah2017TowardsIoT}, as well as the development of new scenarios and business opportunities \cite{Zhang2017TheThings}.

The versatile nature of the bockchain technology is further corroborated by versatile applications in a wide area of domains \cite{Sun2016Blockchain-basedCities, Chen2017PersonalChallenging, Fukumitsu2017ABlockchain}.

The blockchain technology can also be applied in higher education. Several higher educations institutions have employed the blockchain technology to design different solutions and approaches related to higher education. The majority of solutions use the Bitcoin blockchain \cite{InitiativeDigitalProject, UniversidadPlata}. Nazaré et al. have proposed a platform for creating, sharing, and verifying blockchain-based educational certificates within the scope of the Digital Certificates Project. This incubation project is based on the Bitcoin blockchain and is lead by the Media Lab Learning Initiative at the Massachusetts Institute of Technology (MIT).
This approach addressed the issues of digitizing academic certificates and does not investigate the possibility of the blockchain to be used in a global higher education credit and grading platform.

The National University of La Plata (UNLP) has started developing a framework for a blockchain-based verification of academic achievement \cite{InitiativeDigitalProject, Third2016BlockchainsPaper, Bond2015BlockchainCase}, but no further details have been revealed until today \cite{UniversidadPlata}. The same approach was also adopted by the Argentinian College CESYT \cite{Amati2015}. Both solutions use blockchain technology and cryptography (i.e. digital signature, time stamps, etc.) to issue diplomas for students. However, their approach does not address the issue of obtained credits for completed academic achievements. The approach is focused only on issuing diplomas (degrees) using the bitcoin blockchain. 

In 2016, the Parisian Leonardo da Vinci Engineering School (ESILV) announced it would certify diplomas on a bitcoin blockchain \cite{Das2016a}. They have partnered with the French Bitcoin startup Paymium, but no further details or a prototype have been published so far. 
There are also other higher education institutions, which have or intend to use the blockchain technology. In 2015, a software engineering school in San Francisco, the Holberton School, announced using the technology to help employers verify academic credentials \cite{Coleman2015EngineeringCryptoCoinsNews}.

Most of the aforementioned projects in the higher education domain rely on closed concepts or ideas and often do not discuss details or even remain on an idea level. Some of the related projects are offered exclusively to a closed circle of entities. 

On the contrary, the idea presented in this paper relies on open-source technologies (i.e. open-source public code of the implementation) and it aims to incorporate global stakeholders into the EduCTX initiative, and is therefore open for participation and inclusion of any HEI through a publicly available platform and web presence. The presented platform is based on the ARK blockchain technology and a prototype implementation is available via the Github software development platform. 
The proposed EduCTX blockchain platform is thus the basis of the EduCTX initiative, opened globally to all HEIs for building an efficient, simplified, ubiquitous solution for student's credit assignment, while eliminating language and administrative obstacles.

\section*{Background}
\label{sec_background}

\subsection*{European Credit Transfer and Accumulation System}
\label{sec_ects}

The European Credit Transfer and Accumulation System (ECTS) is a framework for the higher education grading system developed by the European Commission and agreed to by the EU member states. It was established in 1989 within the Erasmus (students exchange) program. The objective of this learner-centered system is to facilitate planning, delivery and evaluation of study programs as well as to facilitate student mobility by recognizing prior learning achievements, qualifications, experience and learning periods \cite{Commission2015ECTSGuide}. ECTS credits express the volume of learning, which is based on the defined learning results and the workload that is associated with learning. Learning results are expressed as the knowledge of an individual (what he/she knows, understands and is able to do), while the workload is the estimated time an individual needs to complete all learning activities. Learning results and the associated workload of a full-time academic year is assessed with 60 ECTS credits. Credits are expressed in whole numbers. Considering the expected learning results and the estimated workload, different amounts of credits are assigned to different educational components (courses). One credit corresponds to 25 to 30 hours of work \cite{Commission2015ECTSGuide}.    

\subsection*{Blockchain - Distributed Ledger Technology}
\label{sec_blockchain_types}

A Blockchain can be referred to as a distributed database, that chronologically stores a chain of data packed into sealed blocks \cite{Alex2017OmniPHR:Records} in a secure and immutable manner. The chain of blocks, also called a ledger, is constantly growing, thus new blocks are being appended to the end of the ledger, whereby each new block holds a reference (more precisely a hash value) to the content of the previous block \cite{Sleiman2015BitcoinSystem}. The content of the blocks can be predefined or randomly generated by the users of the blockchain. Nevertheless, the data is structured into so-called transactions according to the predefined structure of the blockchain and cryptographically sealed \cite{Aste2017BlockchainIndustry}. The public key encryption mechanism is used to ensure the security, and thus consistency, irreversibility and non-repudiability of the distributed ledger content \cite{2003PublicCryptography, Zheng2017AnTrends}.
Prior to the sealing of a data block, a cryptographic one way hash function is applied (e.g. SHA256), ensuring anonymity, immutability and compactness of the block \cite{NIST2012SecurePUBLICATION}.   
The ledger and its contents is replicated and synchronized across multiple peers in a P2P network, therefore becoming a distributed ledger. Although the blockchain is a part of the distributed ledger technologies (DLT), not all DLT employ a chain of blocks. We will henceforth refer to the above-mentioned description of the technology as the blockchain \cite{Aste2017BlockchainIndustry}. 

There are three main types of blockchains: (1) public - permissionless, (2) private - permissioned, and (3) consortium blockchains. The permissionless blockchain type emphasizes the public part, hence all the blockchain data is accessible and visible to the public. However, some parts of the blockchain could be encrypted in order to preserve a participant's anonymity \cite{Zheng2017AnTrends}. Furthermore, in these public blockchain types everyone can join the network as a network node. Examples of such a blockchain are the Bitcoin and the Ethereum blockchains. In contrary, a private blockchain enables only chosen nodes to join the network, thus being regarded as a form of a distributed but still centralized network \cite{Zheng2017AnTrends}. The consortium blockchain is a mixture of the two and enables only a selected group of nodes to participate in the distributed consensus process \cite{Zheng2017AnTrends}.

\subsubsection*{Distributed consensus}
\label{sec_consensus}

Since the blockchain is a distributed ledger-like database that rests on a P2P network, each peer of the network holds a replication of the confirmed ledger state and a pool of unconfirmed data which needs to be packed into blocks and added to the ledger. In order for the blockchain network to remain functional, the peers need to agree on a certain state of the ledger content and on a way for packing data into blocks. This is reached by a distributed consensus protocol, which validates the chronological order of data generated \cite{Zheng2017AnTrends}. The distributed consensus protocol ensures that a quorum of blockchain network peers agree on the precise state of the shared ledger, thus the order in which new blocks are added to the ledger \cite{Aste2017BlockchainIndustry}. Some of the used distributed consensus algorithms are \textbf{proof-of-work} (PoW), \textbf{proof-of-stake} (PoS), \textbf{delegated-proof-of-stake} (DPoS), proof-of-importance, proof-of-activity, proof-of-burn, proof-of-deposit, etc. \cite{Borge2017Proof-of-Personhood:Cryptocurrencies, Zheng2017AnTrends, Aste2017BlockchainIndustry}. Next we present in detail PoW and PoS, the most commonly used approaches to reach a consensus in a blockchain. 

The basic idea of distributed consensus protocols is to agree on a peer who will prepare and seal the newest block with still unconfirmed and unpacked data. There are several ways on how to decide or select that peer. The simplest one is to determine it randomly, but such an approach is not effective in terms of network longevity and can even be dangerous for the network, since peers could decide to attack the whole network \cite{Zheng2017AnTrends}. The idea behind PoW, PoS and others is the fact that the chosen peer has to contribute something valuable, which will lead to competition and the best will get a reward, thus mitigating the chances of a possible attack.

The PoW is a consensus protocol used in the Bitcoin network and it uses computing power as a mechanism to determine the chosen peer \cite{Nakamoto2008Bitcoin:System, Zheng2017AnTrends, Yli-Huumo2016WhereReview}. The competition between peers is based on hashing unconfirmed transactions - data, therefore a peer's chance of being chosen is in proportion to its computational power. Each time a peer wins, it gets a reward, which in terms of the Bitcoin network are 12.5 newly generated bitcoins added to its account \cite{Aste2017BlockchainIndustry}. The hashing competition (i.e. mining) is based on the calculation of a block, containing unconfirmed transactions and a random nonce. It is required that the outcome of the hash is equal to a predefined value. In case a peer (i.e. miner) reaches the required value, it broadcasts the newly generated block to other peers, whereby it gets validated and in case of its correctness is appended to the replicas of the distributed ledger by all peers \cite{Zheng2017AnTrends}. 

The PoS consensus protocol is based on the stake of the network value a peer has under control (i.e. their asset). In this case, a peer's chance of being chosen to be the new signer of a block is in proportion to its wealth i.e. stake. Practically, this is executed in the form of a peer depositing a predefined minimum number of its wealth, thus buying a ticket in order to be in the group of peers which will be chosen in a deterministic pseudo-random way as the new block signer. Since the competition in the PoS is not based on the computational power of peers, there is no energy consumption as in the case of PoW, but such an approach is like a shareholder corporation, where the rich have an advantage \cite{Borge2017Proof-of-Personhood:Cryptocurrencies}. Moreover, it is also less likely that a peer will attack the network, since in this case it would attack its own assets \cite{Zheng2017AnTrends}. There are multiple versions of the PoS consensus protocol, whereby each introduces a different approach on how to choose the signer in order to ensure fairness. One of these versions is the DPoS, whereby the difference between a regular PoS system and a DPoS system can be compared to the difference between direct democracy and representative democracy, since stakeholders vote in order to decide the signer, i.e. the delegate \cite{Zheng2017AnTrends}.

\subsubsection*{Blockchain network node}
\label{sec_node}

In order to have a fully functional blockchain network, a set of network nodes is required, which are thus the backbone of the blockchain. Since the blockchain network is kind of P2P network, a node can be regarded as a peer when it starts to connect and communicate with other nodes in the network, thus the appropriate name would be a peer-node. For the sake of convenience, we will henceforth denote it as a node. Technically speaking, a blockchain node is any computer that has the core blockchain client installed and operates a full copy of the blockchain ledger \cite{Greenspan2016BlockchainsDatabases, Zheng2017AnTrends, Lewis2015ATechnology}. 

When the users of a specific blockchain engage with the blockchain, they actually connect to the network through a node \cite{Greenspan2016BlockchainsDatabases}. The so-called miners of the PoW concept are a subset of nodes, since all miners must also operate a fully functional node, thus each miner is a node, but not every node is also a miner. This fact is also evident in terms of other blockchain versions where other distributed consensus types are introduced and no mining is required, e.g. PoS \cite{Yli-Huumo2016WhereReview}. Therefore, we can state that nodes determine the distributed consensus, thus agreeing on a specific rule and that the blockchain acts as a consensus mechanism to ensure that the nodes stay in sync \cite{Lewis2015ATechnology}. Moreover, in the PoW type blockchain, a simple node does not get any reward as a miner does, hence the only benefit for running a node is to help protecting the network \cite{Lewis2015ATechnology}. The basic tasks of a blockchain node are: (1) connecting to the blockchain network, (2) storing an up-to-date ledger, (3) listening to transactions, (4) passing on valid transactions into the network, (5) listening for newly sealed blocks, (6) validating newly sealed blocks - confirming transactions, and (7) creating and passing on new blocks \cite{Greenspan2016BlockchainsDatabases, Zheng2017AnTrends, Lewis2015ATechnology}.

\subsection*{Multisignature protocol}
\label{sec_multisig}

A multisignature protocol is a well known concept in the public key cryptography world \cite{2003PublicCryptography, Gilboa:1999:TPR:646764.703977}. It enables multiple parties to jointly digitally sign an agreed message, each with their own private key. Such an option is desirable in cases when multiple parties have to uniformly agree, as in the case of joint bank account. Such a bank account is for example an International Bank Account Number (IBAN), which in the case of an incoming transaction, does not require any action from the account holders and even more, it hides the identities of the account holders. On the contrary, when an outgoing transaction has to be conducted, each account holder has to give their approval prior to the transaction being processed by the bank. 

Such a concept is already a common practice in the cryptocurrency world, whereby M-to-N blockchain wallets can be created \cite{7846935}. Here M denotes the minimum number of required signers of a transaction and N denotes the full number of possible addresses (account holders). An example would be a 2-3 multisignature address, which consists of three parties (their public keys), and at least two of those have to sign a transaction from this multisignature address in order to be processed. In the cryptocurrency world, using such an approach in order to conduct transactions is referred to as pay-to-script-hash (P2SH), in contrast to the usual pay-to-public-key-hash (P2PKH) transactions, which facilitate non multisignature addresses. 

\section*{The Proposed EduCTX Platform}
\label{sec_prop_concept}

\begin{figure}[!h]
 \centering
\includegraphics[width=\textwidth]{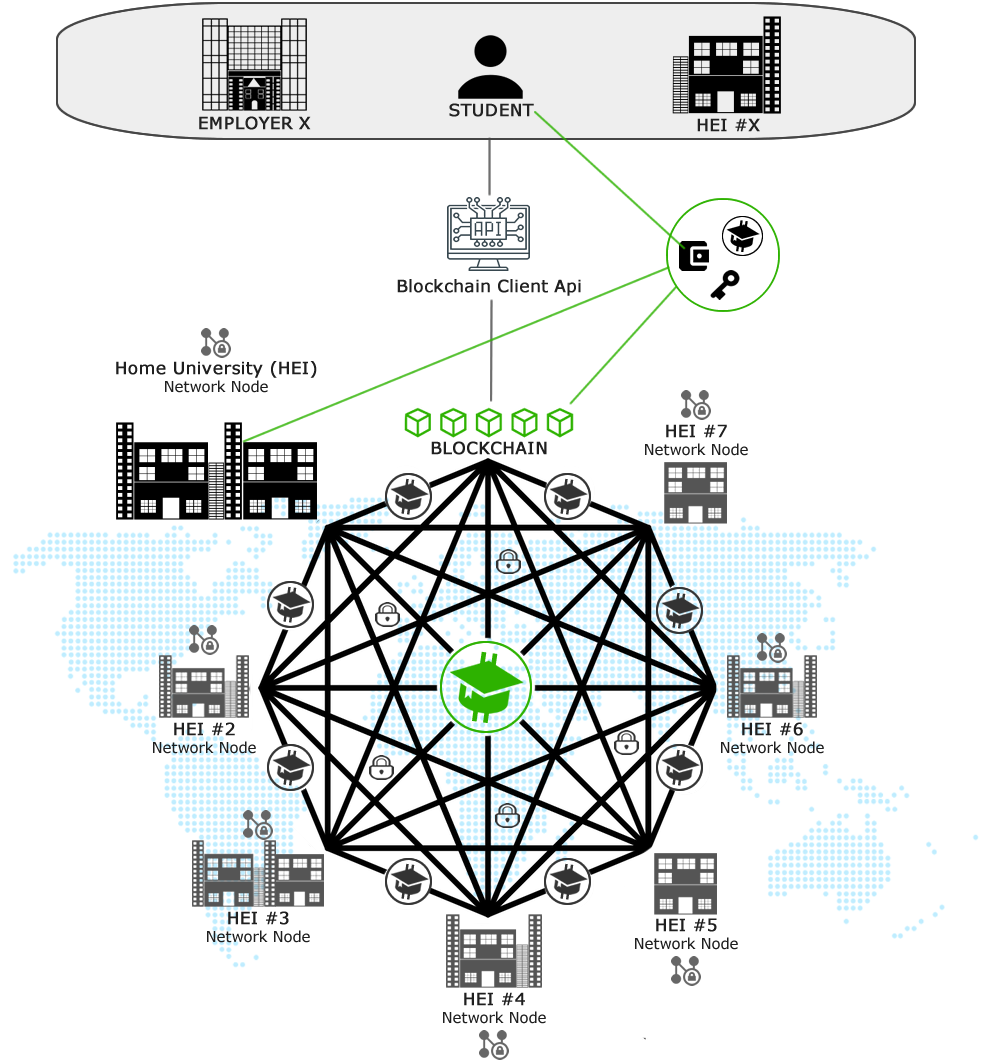}
\caption{A high-level depiction of the proposed EduCTX platform.}
\label{eductx_img}
\end{figure}

This section outlines the proposed platform EduCTX, a blockchain-based higher education credit and grading platform. An abstract depiction of the platform on a higher level is presented in Fig.~\ref{eductx_img}. The EduCTX blockchain platform is envisioned for processing, managing and controlling ECTX tokens as academic credits and resting on a globally distributed P2P network, where peers of the blockchain network are HEI and users of the platform are students and organizations (e.g. companies as potential employers).  

The ECTX tokens represent an equivalent to student's credit value for completed courses, as with the ECTS credits European students gain (see section European Credit Transfer and Accumulation System). Each student will hold a dedicated EduCTX blockchain wallet, where he/she will collect ECTX tokens, i.e. the value of credits assigned by the HEI for his/her completed courses. Every time a student completes a course, his/her home HEI will transfer the appropriate number of ECTX tokens to his/her blockchain address. The transfer information is stored on the blockchain, where the following data is stored: (1) the sender is identified as the related HEI with its official name, (2) the receiver - student is anonymously presented, (3) token - course credit value, and (4) course identification. Furthermore, using his/her blockchain address, the student as the receiver of ECTX tokens, will be able to globally prove his/her completed courses, without any administrative, script or language obstacles by simply presenting his/her blockchain address. For the sake of security, students will be assigned a 2-2 multisignature address by their home HEI, thus they will not be able to transfer any of the gained ECTX tokens to other addresses (see section Multisignature protocol). The process of assigning students with ECTX tokens and their ability to prove the possession of those will be handled through a simple to use EduCTX blockchain API client, thus making the use of the EduCTX platform as intuitive as possible.  

Any accredited HEI and their members will be able to join the network. While joining the network the HEI will have to set-up a network node (see the section Blockchain network node) in order to maintain a global infrastructure and a secure network. A fully functional node broadcasts messages across the network, which is the first step in the transaction process that results in a block confirmation, thus confirmation of ECTX credits transfers for completed courses to students. The HEI node will also have the core EduCTX blockchain client on its server instance with the replication of the complete blockchain ledger. This adds to the security, since the more nodes there are, the more secure the network is.      

The HEI and thus nodes, will not have to mine transactions, since the EduCTX blockchain platform will be based on DPoS consensus protocol (see the section Distributed consensus). Therefore no computing power is needed by the HEI node. Such an approach is also appropriate from the security aspect for the EduCTX network, since random peers cannot join the network and generate new ECTX tokens by mining them. As such, the EduCTX blockchain can be viewed as a consortium version of a blockchain (see the section Blockchain - Distributed Ledger Technology). Each new HEI that joins the network and is reviewed by other member HEIs, will be assigned with ECTX tokens and asked to set up a network node. Since we propose a DPoS distributed consensus version of the blockchain, each HEI member will be able to register as a delegate in the EduCTX blockchain platform and the EduCTX HEI community will vote for a delegate which in turn will confirm transactions and seal blocks. This implies that the community will vote for that HEI which will be the most affirmative and continuous in its work. In order to insure a permissioned version of the blockchain platform, and a democratic and non-profit community, we plan to lower the forging reward to zero.

In the following subsections we describe in detail the four important scenarios, which will play out in the EduCTX. Each scenario is backed up by a Business Process Management (BPM) diagrammatic representation.

\subsection*{HEI joining the EduCTX network}
\label{sec_scenario_1}

\begin{figure}[!h]
 \centering
\includegraphics[width=\textwidth]{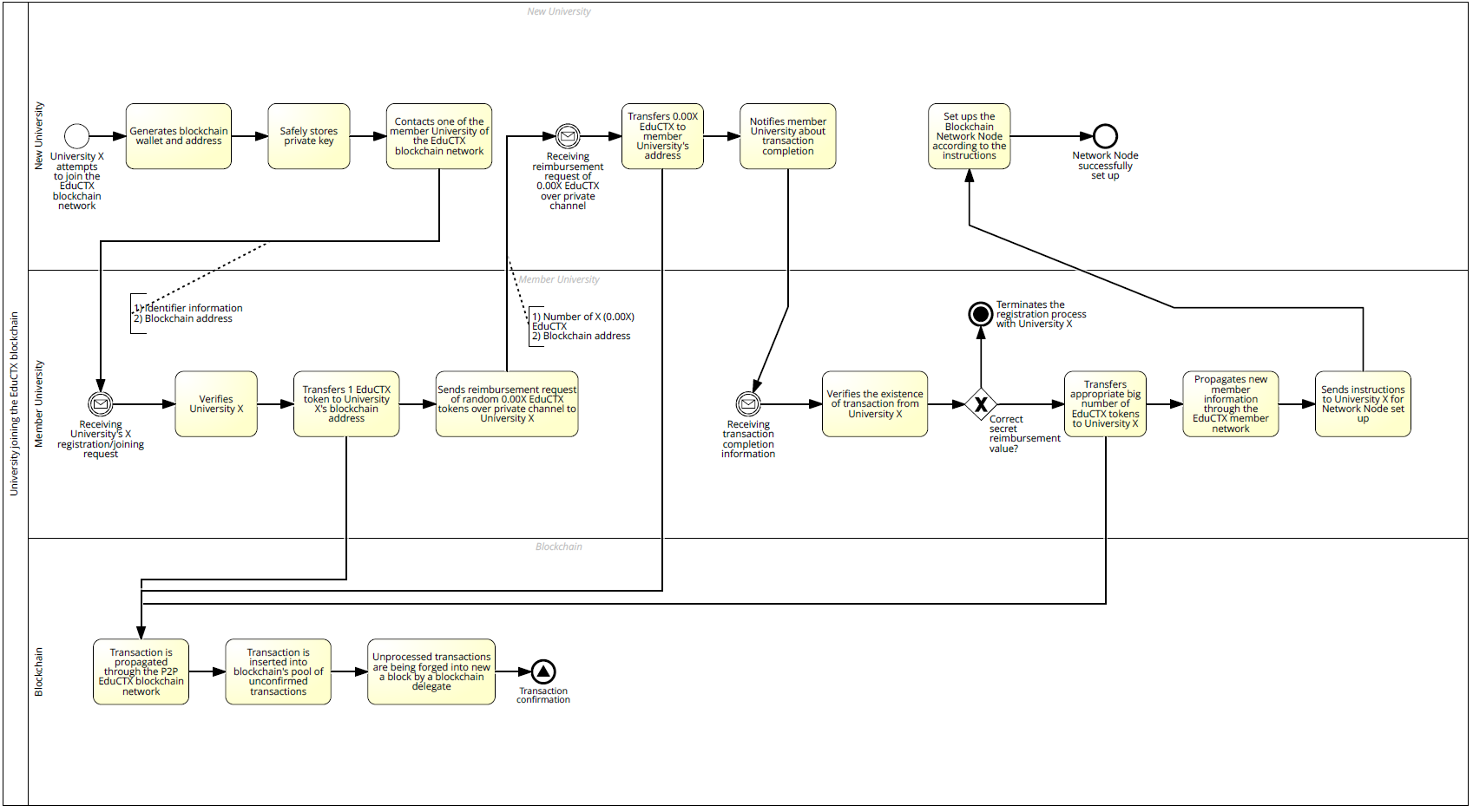}
\caption{A process model of a new higher education institution joining the EduCTX blockchain network.}
\label{uni_join}
\end{figure}

A new HEI (hereinafter referred to as newHEI) attempts to join the EduCTX blockchain network using our API to generate its blockchain wallet and address containing public and private keys. The newHEI should safely store its obtained private key. After generating the address, it contacts one of the existing HEIs, members of the EduCTX blockchain network (hereinafter referred to as memHEI). The memHEI receives a newHEI registration (joining) request. It firstly verifies the newHEI's official information and afterwards transfers 1 ECTX token to the newHEI's blockchain address. The transaction is then processed through the blockchain network (for details refer to Fig.~\ref{uni_join}). When the transaction is confirmed, the memHEI sends a reimbursement request of random 0.00X ECTX
tokens over a private channel to newHEI. It includes (1) the number of X (0.00X) EduCTX (e.g. 235781 - meaning 0.00235781 EduCTX) and (2) the blockchain address. When newHEI receives the reimbursement request over a private channel, it transfers 0.00X EduCTX to memHEI's address (the transaction is processed through the blockchain network). Afterwards, newHEI notifies memHEI after the transaction is completed. memHEI verifies the existence of the transaction from newHEI. If the secret reimbursement value is incorrect, memHEI terminates the registration process. Otherwise, it transfers the appropriate number of ECTX tokens to newHEI (the transaction is being processed through the blockchain network). memHEI propagates the information on newHEI through the EduCTX member network and sends instructions to newHEI for a Network Node set up. newHEI set ups the blockchain network node according to the instructions. After the network node is successfully set up, the process of newHEI joining the EduCTX blockchain is completed. Details about the process model for the described scenario are given in Fig.~\ref{uni_join} in the BPM diagrammatic representation.

\subsection*{Student's registration}
\label{sec_scenario_2}

\begin{figure}[!h]
 \centering
\includegraphics[width=\textwidth]{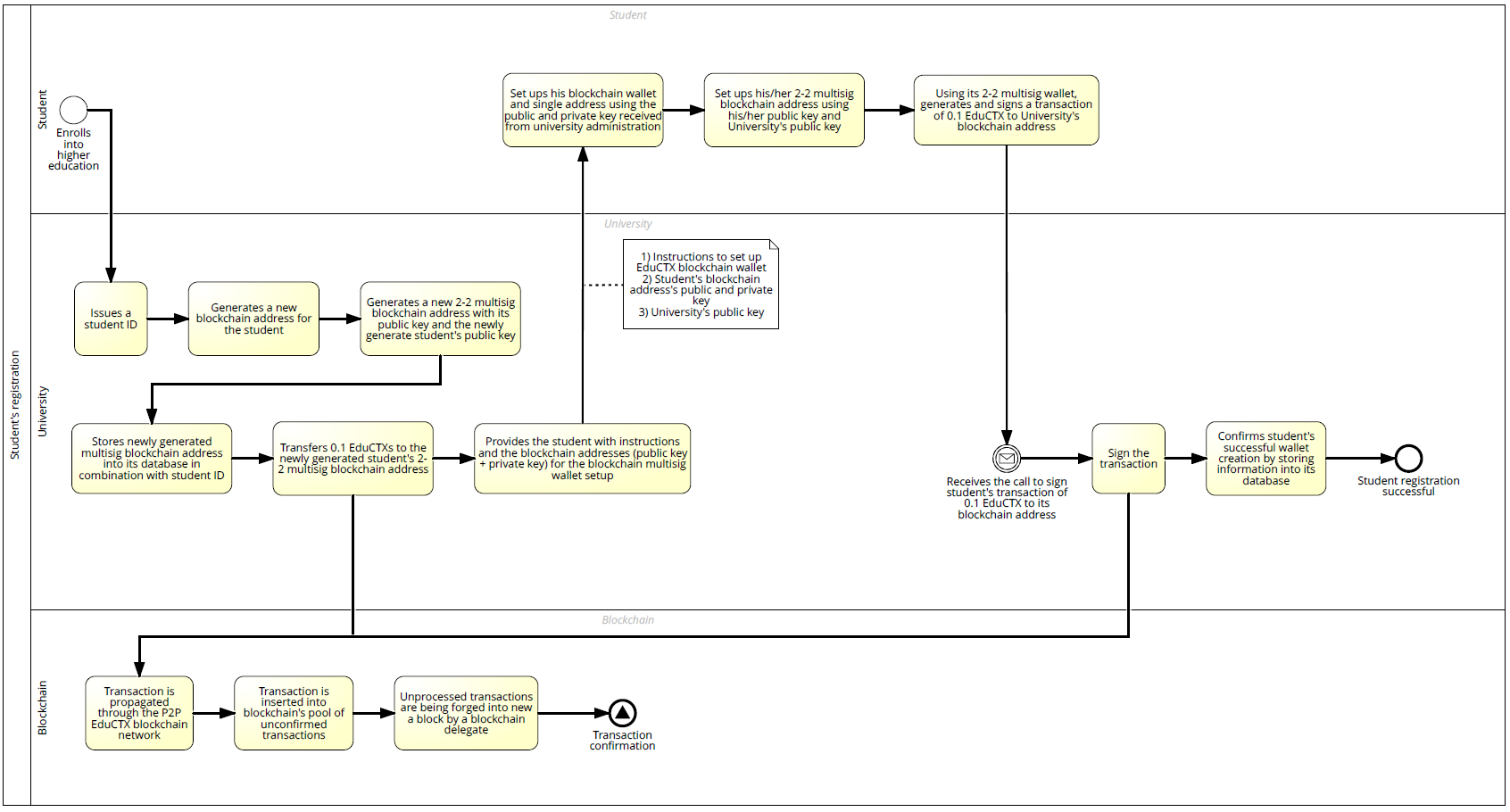}
\caption{A process model of a student's registration into the EduCTX blockchain network.}
\label{student_reg}
\end{figure}

When a student enrolls in the HEI (member of the EduCTX blockchain network), it issues a student ID and generates a new blockchain address for the student, containing a public and private key. Additionally, the HEI generates a new 2-2 multisignature blockchain address with its public key and newly generated student's public key. This multisignature address in combination with student ID is stored in the HEI's database. The HEI transfers 0.1 ECTX token to the student's 2-2 multisignature blockchain address and over a private channel provides the student with the information needed for the blockchain multisignature wallet setup. The information provided includes (1) instructions to set up an EduCTX blockchain wallet, (2) the student's blockchain address containing public and private keys, (3) the HEI's public key and (4) the redeem script. With the received information the student set ups his/her blockchain wallet and a single address using the public and private keys received from the HEI administration. He/she also sets up a 2-2 multisignature blockchain address using his/her public key and HEI's public key. The wallet data should be safely stored. Using his/her 2-2 multisignature wallet, the student generates and signs a transaction of 0.1 ECTX token to the HEI's blockchain address. Afterwards, the HEI signs the transaction using its private key. The transaction is processed through the blockchain network. When the transaction is confirmed, the HEI stores the information in its database, confirming the student's successful wallet creation. Fig.~\ref{student_reg} depicts a process model for the described scenario using the BPM diagrammatic representation.

\subsection*{Student's course completion}
\label{sec_scenario_3}

\begin{figure}[!h]
 \centering
\includegraphics[width=\textwidth]{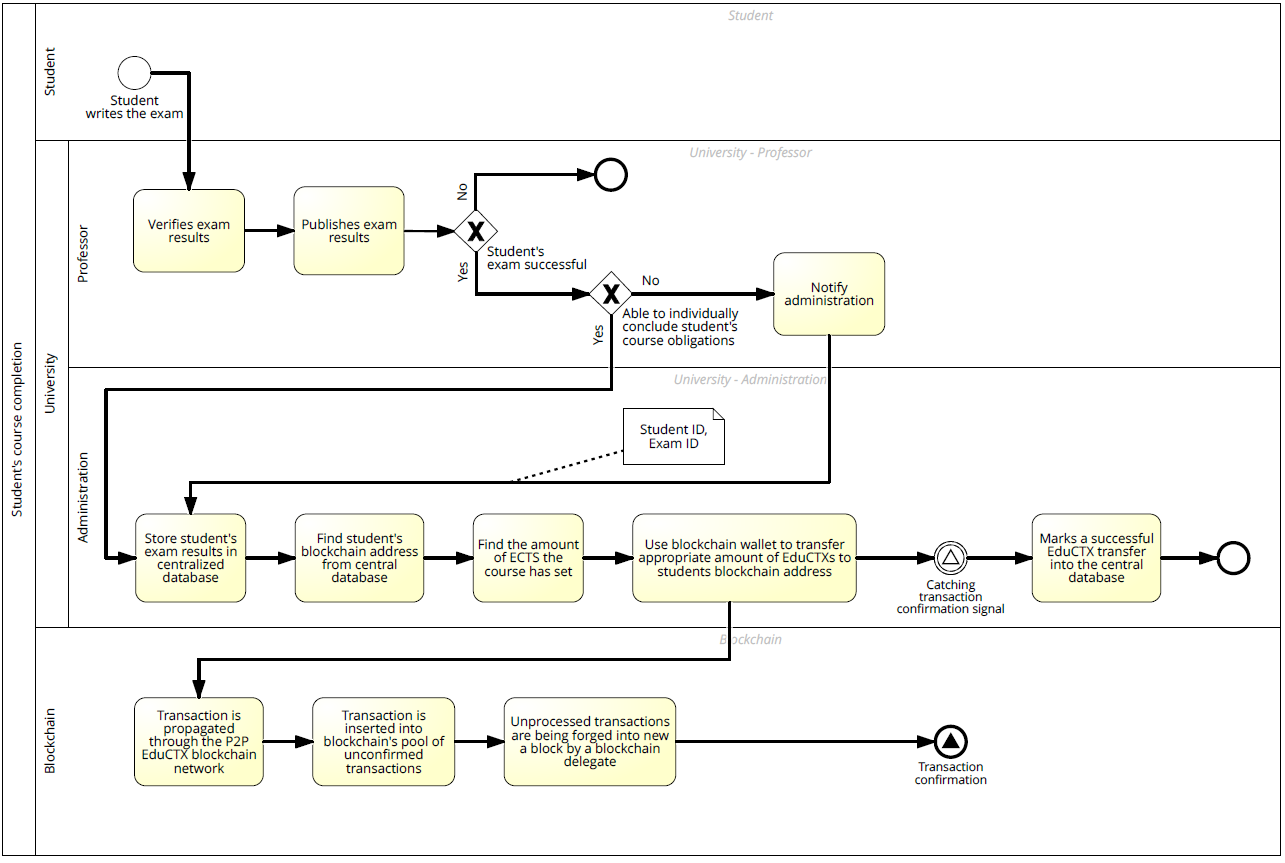}
\caption{A process model of a student's course obligation completion registered in the EduCTX blockchain network.}
\label{student_course}
\end{figure}

After a student takes an exam, the professor needs to verify the results. He/she publishes the exam results. If the student has been successful and if the professor is able to individually register completion of student's course obligations, then the results are stored in the centralized database. Otherwise the professor notifies the administration office to perform the procedure needed to register the status of the student's obligations. The results can be simultaneously stored in the centralized database in case the HEI needs to keep a parallel system conditioned by national law regulations and also as a backup until the EduCTX blockchain system does not prove itself as fully implemented und running. The professor or the administration office finds the student's blockchain address in the central database, finds the amount of ECTS the course has set and uses the blockchain wallet to transfer the appropriate amount of ECTX tokens to the student's 2-2 multisignature blockchain address. The transaction is processed through the blockchain network. When the transaction is confirmed, the professor or administration office records the successful ECTX tokens transfer into the central database. A process model for the described scenario is depicted in Fig.~\ref{student_course}, using the BPM diagrammatic representation.

\subsection*{Organization verifies student's credit record}
\label{sec_scenario_4}

\begin{figure}[!h]
 \centering
\includegraphics[width=\textwidth]{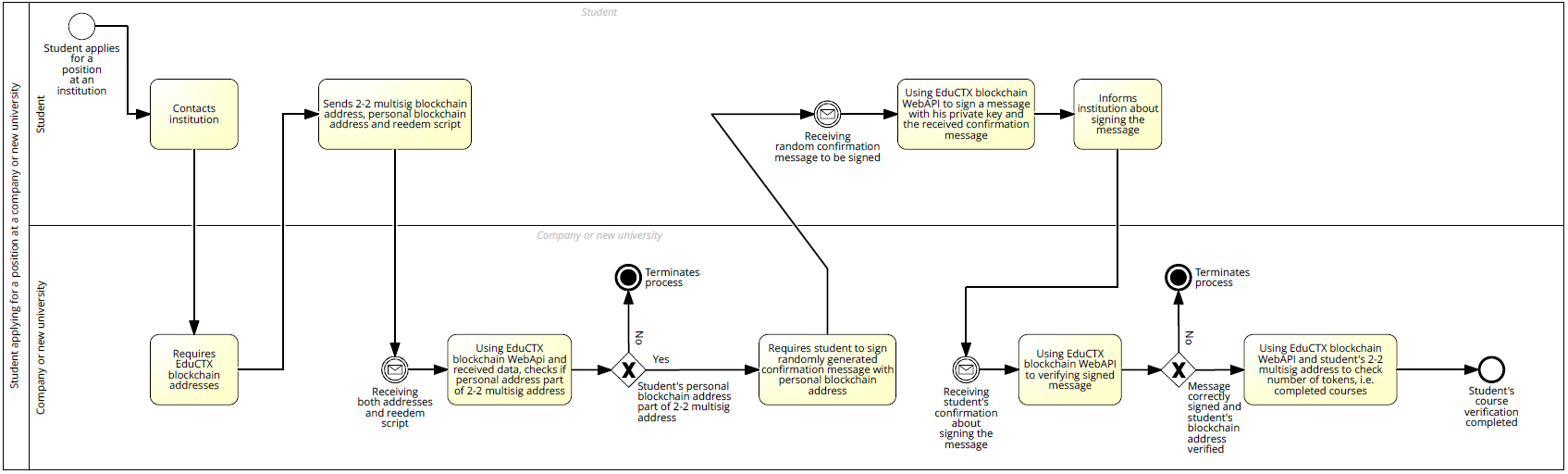}
\caption{A process model of an organization verifying a student's credit record.}
\label{student_joining_company}
\end{figure}

When an organization (e.g. employer, university, etc.) wants to verify the student's course obligation completion, the student has to send his/her blockchain address, his/her 2-2 multisignature blockchain address and redeem script to the verifier - organization. The organization checks the redeem script to verify a student's address and 2-2 multisignature address. Using the blockchain web API to access blockchain data, the organization checks the amount of ECTX tokens in the 2-2 multisignature address, which represents the student's academic credit achievements. Afterwards, via a private channel, the organization requires that the student signs a message (e.g. "XYZ") with his/her address in order to verify his/her identity. When the student, using the blockchain web API, signs the message with his/her address and private key, he/she notifies the organization, who checks the signed message. If the signed message is validated, the organization can trust that the presented blockchain address and its ECTS tokens value are indeed those of the student. A process model for the described scenario is depicted in Fig.~\ref{student_joining_company}, using the BPM diagrammatic representation.

\clearpage

\section*{Prototype implementation}
\label{sec_prototype}
We selected the ARK Blockchain \cite{2016All-in-OneSolutions} as the underlying technology of our EduCTX platform. ARK is not only a cryptocurrency, but is also an ecosystem meant for blockchain mass adoption. By building the EduCTX platform on top of a highly secure and fast ARK core blockchain, integrating key decentralized technologies, the platform becomes a user-university-friendly ecosystem to increase the adoption of blockchain technology as a whole. The main reasons for selecting the ARK technology as a code base are its flexibility and open-sourceness, and the overall availability of a client API implementations. At the time of writing, ARK provides more than 12 different programming languages of client implementations, thus enabling other actors (HEIs, students, employers) to join the platform in the programming language of their choice.

\subsection*{EduCTX ecosystem building blocks}
\label{sec_eductxplatform}
ARK is built in order to utilize a consensus mechanism known as DPoS. While DPoS (see section Distributed consensus) has been known for quite some time, it has not received the same level of exposure as PoW and PoS \cite{2017ARKAdoption}. The DPoS consensus model has several major differences when compared to traditional PoW and PoS models. DPoS can be viewed as a representative democracy where individual users (in our use case HEIs) utilize their stakes (shares, trust) in order to nominate delegates (other HEIs) to join the EduCTX network of trustworthy nodes (delegates). Delegates are  responsible for validating transactions and securing the network. Should delegates (EduCTX member HEIs) perform their duties poorly, or use authority in a manner not representing the initial agreement, votes may be rescinded and assigned to a new or existing representative - all in the name of ensuring the security and trust of the established network.

ARK was also selected for its flexibility. It is a permissionless blockchain, meaning that anyone can join. We adapted the ARK blockchain to a consortium type blockchain by changing the network DPoS consensus parameters. New network members can join only based upon pre-agreement, by proving their identity within current existing centralized systems (e.g. a signed statement from the HEI dean).  

In terms of the security and validity of the EduCTX records on the blockchain we defined several rules, to ensure the safety and validity of student course completion records. 
\begin{itemize}
    \item Every student is anonymous. A student is presented with his/her unique blockchain address and this address will store and receive ECTX tokens, thus ECTS-like credit value, that will confirm his/her completion of various courses.
    \item Students cannot send received ECTX tokens, thus ECTS-like credits, to another address. It is a 2-2 multisignature address that would require a second signature in order to enable outgoing transactions of ECTX tokens (see Section Multisignature protocol). A second signature is required by an authorized entity, that issued the student address, i.e. his/her home HEI.
    \item Although a student is fully anonymous, we implemented a messages signature system, that can be used to prove the identify (ownership) of the ECTX tokens.
\end{itemize}

A public EduCTX blockchain explorer is already available at http://eductx.um.si. The EduCTX block explorer is an online blockchain browser, which displays the overall state of the EduCTX network (delegate nodes, availability, contents of individual blocks, transactions, transaction histories and EduCTX balances of addresses).

\subsection*{Joining EduCTX ecosystem}
The EduCTX code is published on GitHub \cite{eductxgithub} and licensed under the MIT License \cite{TheLicense}. Anyone can review and adapt the code to its own needs, thus also contributing to the whole EduCTX ecosystem, by sharing new features to all users - what is also the underlying concept of open-source projects and our EduCTX initiative. We encourage and invite other HEIs to join, contribute and expand the EduCTX concept further.

We designed the EduCTX in a modular manner (based on the building blocks of the ARK blockchain ecosystem). EduCTX can be seamlessly integrated with existing HEI's information systems. The blockchain end points are REST APIs that can be consumed by already published and available client APIs (see EduCTX github for more details). The EduCTX platform is meant to co-exist with existing HEI information systems in various countries. It is not intended to replace them, but co-exist and provide efficient, trustworthy and above all transparent and immediate proof of course completion. 

\begin{figure}[!h]
 \centering
\includegraphics[width=\textwidth]{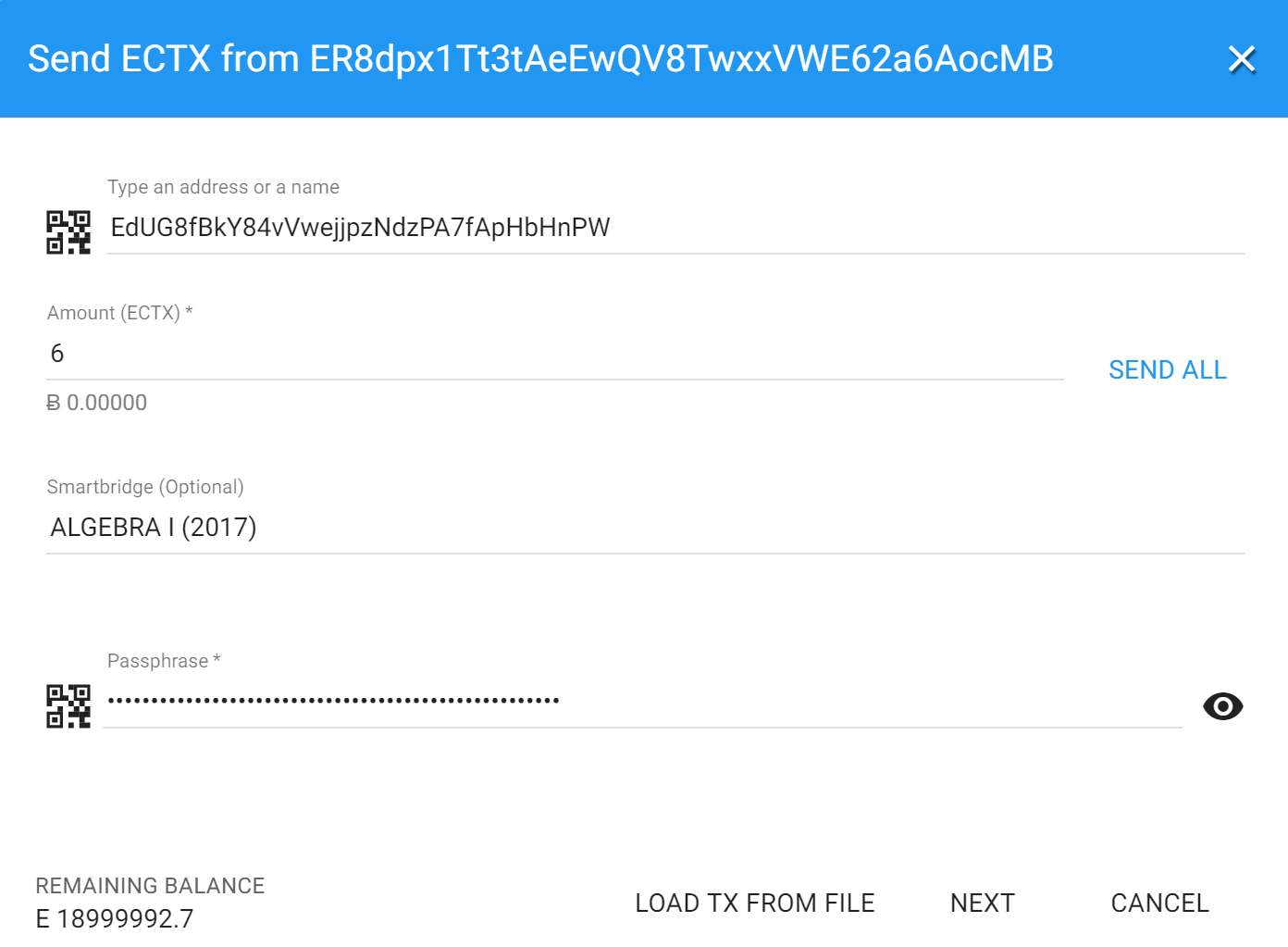}
\caption{Professor assigning credits using the ECTX client wallet.}
\label{prototype}
\end{figure}

As a first user-interface to the EduCTX ecosystem, the ECTX client (wallet) application is available (see Fig.~\ref{prototype}). Using the wallet, universities can register new users (addresses for students), transfer EduCTX credits to students (represented by an address) and perform an initial registration of the applying HEI (delegate registration). The wallet provides a secure interface for the blockchain ecosystem by leveraging the REST API provided by blockchain network nodes. For convenience, the wallet also enables hardware security - by integrating the Ledger Nano S hardware (HW) wallet. Ledger Nano S is a HW wallet, based on robust safety features for storing cryptographic assets and securing digital payments \cite{LedgerWallet}.

In order to join the EduCTX ecosystem, the applying HEI has to provide a technical base (run at least one main network node, deliver a wallet application to its applicants) and prove its identity. The basic steps include: (1) preparing an EduCTX node, (2) joining the EduCTX network, and (3) becoming a permissioned EduCTX member. After careful identity evaluation, the new HEI can be accepted as a legal and trusted representative of the EduCTX network. The new HEI is voted into the delegate network by other trustworthy HEIs (delegate nodes). Detailed instructions are available at the official web page of the EduCTX Initiative, eductx.org.

The EduCTX ecosystem has been designed to run efficiently, securely and non-invasively. Integration with existing information systems will be conducted based on their requirements by using the REST APIs. There are many benefits for the end-users (trustworthy completion records, cross-border course confirmations, fast transactions, employee verification), that will be delivered via existing and new applications based on the REST API. The overall concept of the blockchain ecosystem is to run in the background and provide a global, trustworthy and tamper-proof backbone for a potential academic credit system. 

\section*{Discussion}
\label{sec_results}

The following section outlines the broader context, assumptions, and identifies the limitations and challenges of the proposed blockchain platform.  

With the presented proof of concept and the prototype implementation of the EduCTX platform, we have also introduced the EduCTX initiative, and anticipate that the community around it will grow further. All the identified stakeholders (i.e. HEIs, students, companies) can benefit from a globally trusted, decentralized higher education credit and grading system, which is easy to use and is free from any administrative, script and language obstacles, thus introducing understandable programmatic clarity for a highly frequent process, which is spread throughout the world. 

It should be noted that the EduCTX initiative's intent is not to completely change and transform the current credit and grading systems established in various countries, but to facilitate it by adding transparency and automation in order to optimize administrative processes related to the higher education system on a global scale. This is even more a case considering various legislative rules in different countries and considering various credit and grading systems on a global scale. In fact, due to legal reasons (national or transnational legislation and rules) the coexistence of both systems is a very likely scenario. Therefore, we encourage HEIs to join the EduCTX initiative and use the EduCTX platform simultaneously with their current process management, while also contributing with new ideas, comments and suggestions regarding the EduCTX platform. The openness of both the platform and the concept also enables the advancement and further development of the concept in technological additions or positive modifications to the concept. 

There are several advantages for various stakeholders using a platform such as EduCTX. Principally, the EduCTX platform enables organizations as possible future employers the possibility of checking academic records of potential employees in a transparent way. HEIs get an open, decentralized and transparent way of validating records for students and their obligations. The proposed platform supports the HEIs in their activities related to students and provides the possibility of fraud detection and prevention. Therefore, the need for a more complex checking process regarding a student's academic records can be avoided. Additionally, the students are offered the possibility of transparency and an overview of their academic obligations within the scope of their study programs. A student can instantly check his/her completed course records. An obvious advantage for all the involved stakeholders is also the possibility of an audit trail. In this way fraud can be prevented. 

The proposed initiative and platform will be firstly applied on a smaller number of HEIs and organizations, thus allowing for a first initial phase that can be sanitized of any imperfections in both the technical and conceptual parts. The platform is open to use for any interested party, which we anticipate will attract HEIs around Europe and also globally. 

The research also includes specific presumptions and constraints. The first deployment of the platform will be done by our home HEI, the University of Maribor. The platform needs other HEIs and stakeholders to join the imitative in order to prosper.  The proposed platform also has its specific issues and imperfections. At first only a few nodes will be part of the networks and this can be seen as a security risk. However, we anticipate the number of HEI nodes will grow and in this way the security concern will minimize in a timely manner. Moreover, when an appropriate number of HEI nodes become part of the network the issue will completely disappear. An additional issues are related to the adequate storage and protection of private keys for all participating entities (e.g. student, HEIs). It is presumed that all participants will protect and backup their private keys, as should be expected when dealing with sensitive data. Such practices and aspects are normal in societies which enable a physical person to sign electronic documents with their digital signatures, which in turn are issued by governmental institutions and protected by private keys and passwords \cite{6482707}. Furthermore, it is also required from participants to protect and safeguard their credentials in the physical world, where we expect from the trusted authorities (in this case HEIs) to protect and backup all their official stamps, signatures, etc. 

Nevertheless, a scenario can happen where a student loses his/her private key, thus being unable to prove the possession of a EduCTX blockchain address and ECTX tokens. In this case, the student can personally visit his/her home HEI and request that a new blockchain address be issued to him/her. His/Her home HEI would verify his/her records and again transfer the appropriate number of ECTX tokens to his/her new blockchain address. As an alternative, the student's old address could be annulled by various means. An improvement proposal will be discussed further on within the EduCTX initiative. Moreover, a student's blockchain wallet data (public and private) keys could be stolen by an adversary who could try to impersonate the student and claim his/her academic achievements. This scenario could be avoided by adding an additional level of protection to the student's blockchain wallet data, i.e. encrypting the wallet data with a password or using an already integrated Ledger Nano HW wallet. For additional security, each involved platform stakeholder could be issued with a multi-level multisignature address (e.g. 2-8, whereby all the keys are assigned to the original owner and stored in various secret and safe places), ensuring that an address is functional even if one of the keys is lost.                  
\section*{Conclusion and Future Work}
\label{sec_conclusion_future_work}

EduCTX was proposed as a global blockchain-based higher education credit platform. The proposed platform takes the advantage of the blockchain in order to create a globally trusted higher education credit and grading system. As a proof of concept, we presented a prototype implementation of the EduCTX platform which is based on the open-source Ark blockchain platform. The proposed EduCTX platform addresses a globally unified viewpoint for students and organizations. Students benefit from a single and transparent view of their completed courses, while HEIs have access to up-to-date data regardless of a student's educational origins. Other beneficiaries of the proposed system are potential employers, who can directly validate the information provided by students.       

The proposed solution is based on the distributed P2P network system. It transfers the higher education grading system from the current real-world physical records or traditional digital ones (e.g. databases) to an efficient, simplified, ubiquitous version, based on blockchain technology. It is anticipated that such a system could potentially evolve into a unified, simplified and globally ubiquitous higher education credit and grading system.

We hereby strongly encourage, appeal and invite all HEIs to contact us and thereby join the EduCTX initiative.  

In the future we plan to adapt the EduCTX blockchain so that each course would be assigned with a unique blockchain address and a pool of tokens. After completing the course obligations, students would get tokens from the course address and not directly from the institution. The course address would be a multisignature address between an institution and a professor. 

Currently, the proposed platform is based on the ECTS grading system, but could be potentially extended and adapted to any existing credit or certification system and thus incorporate other aspects of educational certification. 

We further plan to extend our work and the EduCTX platform to be based on smart contracts and an appropriate version of the blockchain technology.

\section*{EduCTX Initiative Invitation}

The EduCTX blockchain platform rests on the idea that all globally distributed HEIs join forces together in order to build an efficient, simplified, ubiquitous solution for a student's credit assignment, while eliminating language and administrative obstacles which arise when dealing with credit transfers and certification on a broader or global scale. Therefore, we invite anyone interested to join the EduCTX initiative (more information is available at eductx.org), so that the EduCTX idea can be further discussed and developed. 
We also encourage all HEIs, especially those already having the ECTS accredited system, to join the EduCTX blockchain network, thus helping spread and add to the security of the network. For further information, please contact us at: eductx@um.si.

\section*{Acknowledgements}
We acknowledge financial support from the Slovenian Research Agency (Research Core Funding No. P2-0057).

\nolinenumbers

%
%
%
\bibliography{mendeley}



\end{document}